\documentclass[Physsubmission, Phys]{SciPost}
\hypersetup{
    colorlinks,
    linkcolor={red!50!black},
    citecolor={blue!50!black},
    urlcolor={blue!80!black}
}

\usepackage[bitstream-charter]{mathdesign}
\DeclareSymbolFont{usualmathcal}{OMS}{cmsy}{m}{n}
\DeclareSymbolFontAlphabet{\mathcal}{usualmathcal}

\begin{document}

\begin{center}{\Large \textbf{
Structure Functions and Parton Densities: a Session Summary\\
}}\end{center}

% TODO: write the author list here. Use initials + surname format.
% Separate subsequent authors by a comma, omit comma at the end of the list.
% Mark the corresponding author with a superscript *.
\begin{center}
Benjamin Nachman\textsuperscript{1$\star$},
Katarzyna Wichmann\textsuperscript{2} and
Pia Zurita\textsuperscript{3}
\end{center}

\begin{center}
{\bf 1} Physics Division, Lawrence Berkeley National Laboratory, Berkeley, CA 94720, USA
\\
{\bf 2} DESY, Notkestrasse 85, 22607 Hamburg, Germany
\\
{\bf 3} Institut fu\"r Theoretische Physik, Universit\"at Regensburg, 93040 Regensburg, Germany
\\
* bpnachman@lbl.gov
\end{center}

\begin{center}
\today
\end{center}

% For convenience during refereeing (optional),
% you can turn on line numbers by uncommenting the next line:
%\linenumbers

\definecolor{palegray}{gray}{0.95}
\begin{center}
\colorbox{palegray}{
  \begin{tabular}{rr}
  \begin{minipage}{0.1\textwidth}
    \includegraphics[width=22mm]{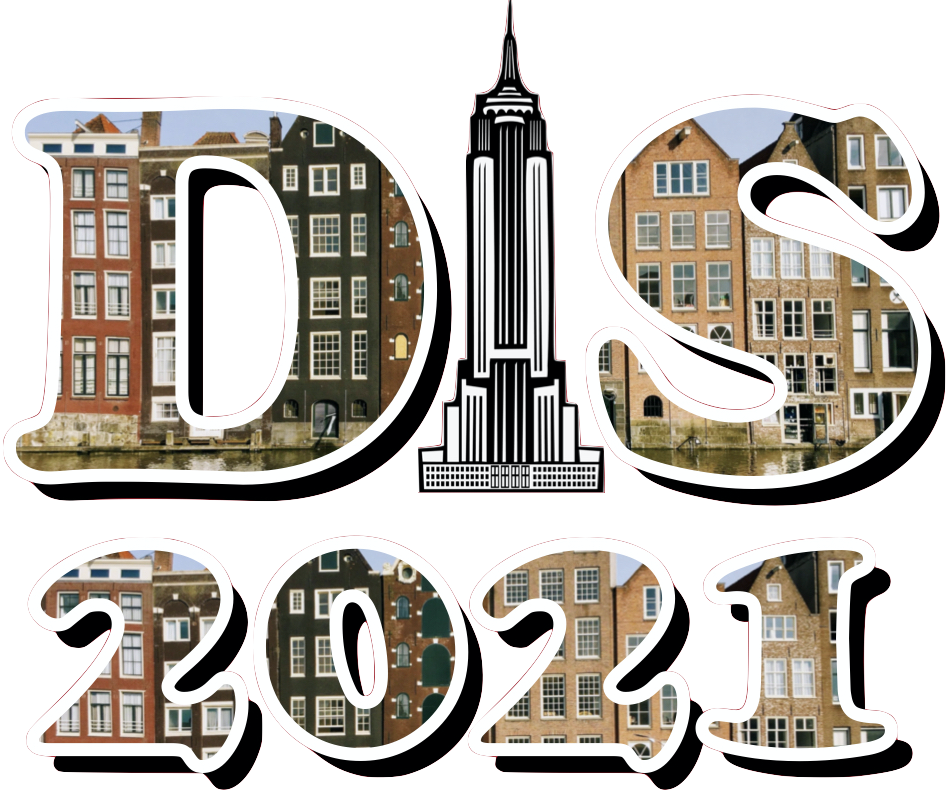}
  \end{minipage}
  &
  \begin{minipage}{0.75\textwidth}
    \begin{center}
    {\it Proceedings for the XXVIII International Workshop\\ on Deep-Inelastic Scattering and
Related Subjects,}\\
    {\it Stony Brook University, New York, USA, 12-16 April 2021} \\
    \doi{10.21468/SciPostPhysProc.?}\\
    \end{center}
  \end{minipage}
\end{tabular}
}
\end{center}

\section*{Abstract}
{\bf
% TODO: write your abstract here.
Studies of fragmentation and parton density functions are a core component of research in high energy particle and nuclear physics. These quantities are inherently interesting as a probe of the quantum nature of the strong force and are also essential ingredients to additional studies in high energy scattering experiments. These proceedings provide an overview of the state of the art in this area, as presented at the Deep-Inelastic Scattering Conference in the Spring of 2021.
%The abstract is in boldface, and should fit in 8 lines.
%It should be written in a clear and accessible style, emphasizing the context, the problem(s) studied, the methods used, the results obtained, the conclusions reached, and the outlook. You can add a table contents, recommended if your paper is more than 6 pages long.
}

\vspace{10pt}
\noindent\rule{\textwidth}{1pt}
\tableofcontents\thispagestyle{fancy}
\noindent\rule{\textwidth}{1pt}
\vspace{10pt}

%%
%% %Our session: https://indico.bnl.gov/event/9726/sessions/3983/#20210415
%% %Planning google sheets: https://docs.google.com/spreadsheets/d/1fI7bn-KS0xIaJRlc5mAcZGIHUm1oaZ45y-TJxCKWO8A/edit#gid=0
%%

\section{Introduction}
\label{sec:intro}

This document summarises the topics discussed in the parallel session \emph{Structure Functions and Parton Densities}. Therefore, it does not aim at being a detailed description of the presented results, which can be found in the accompanying articles of these proceedings. Instead we would like to provide the readers with a list of the key points, and encourage them to read the corresponding contributions.    

\section{Overview of Parton Density Functions}\label{sec:pdfs}

The luminosity upgrade of the LHC \cite{Schmidt:2016jra} and the upcoming Electron-Ion Collider \cite{AbdulKhalek:2021gbh} will provide measurements of unprecedented precision. In order to have meaningful comparisons between data and theory, observables in the pQCD framework will require a more precise knowledge of the cross-section for both the perturbative and non-perturbative components. The former advances slowly, with higher order calculations requiring the computation of an ever increasing number and types of diagrams, while for the latter several sources of improvements are being explored. Recent results addressing both sides of the picture were shown in this conference.   

For the perturbative side, preliminary results towards the calculation of the next-to-next-to-next-to-next-to leading order ($N^{4}LO$) DIS cross-section and four-loop splitting functions were presented. The method relies on obtaining the Mellin moments by reducing the calculation to a set of master integrals, which in turn can be used to reconstruct the coefficient functions. At the time of the presentation the result of the three loop calculation were yet to be compared with the already known one for finite Mellin momenta \cite{Moch:2004pa,Vogt:2004mw,Ablinger:2014nga,Ablinger:2017tan}, the following step being the four loop. 

More loops in the QCD perturbative expansion also means that, at some point, the magnitude of the QED contribution will be comparable to the one from QCD. A new approach for the treatment of these terms, equally applicable to the extraction of collinear and transverse momentum dependent parton distribution functions, was presented \cite{Liu:2020rvc}. In it, the radiative corrections are resummed into sets of universal lepton and fragmentation/jet distributions. 

As much as higher order calculations are necessary, increasing the precision of the non-perturbative parton distribution functions (PDFs) is equally important. While the computation of PDFs from first principles is experiencing significant improvements in the field of Lattice QCD (see Sec. \ref{sec:lattice} for the latest results), we still rely on global fits to the world data for their extraction. In this session several groups have presented improvements to their analyses. The changes can roughly be split into three categories: adding more data, taking into account nuclear corrections to processes involving heavy nuclei as targets, and considering the nuclear corrections in the case of deuterium targets. The last two points have already been included in previous releases of the PDF fits and will not mentioned in the following.

\subsection{Expanding the fitted data set}\label{subsec:adddata}

One of the paths to better constrain the PDFs is the inclusion of more data, which can provide constraints in previously unexplored regions and/or, if the novel data have a higher precision, help reduce the uncertainties. The CTEQ-TEA group presented a comparison of their CT18 NNLO \cite{Hou:2019efy} and accompanying alternative analyses that include the most recent LHC data, with emphasis on dealing with tensions among data sets and properly characterising the goodness of the fit \cite{Kovarik:2019xvh}. 

In the same line, the CJ group introduced the efforts towards the update of their PDF set CJ15 \cite{Accardi:2016qay}. The ongoing study incorporates data from DIS at JLAB at both $6$ and $12$ GeV, electroweak boson production at LHC and RHIC, Drell-Yan at RHIC, and preliminary Drell-Yan data from SeaQuest. These last two have a significant impact in the up and down sea quarks distributions. The RHIC data is compatible within uncertainties with the Drell-Yan measurement of E866 in the $x<0.3$ region. The preliminary SeaQuest data and the one from E866 do not overlap in the high $x$ region, which translates to quite different behaviours of the $\bar{d}/\bar{u}$ ratio in the CJ analysis.

Adding new data is also the basis for the update of the MSHT (formerly MMHT, formerly MRST) group. The MSHT20 set \cite{Bailey:2020ooq} including the final Tevatron data, the combined HERA data from DIS, and some of the vector boson production, inclusive jets and top quark measured at the LHC. A fraction of the LHC data has been shown to have significant tension among them and thus were not considered. The fit was done fully at NNLO (if possible) with NLO QED corrections whenever required, and using more flexible functional forms for the PDFs, which increased the number of parameters to determine but reduced tensions between data sets. The new parametrisation coupled with the extended data set resulted in the reduction of the uncertainty bands of all species, and changes in the shape of the down valence and strange distributions, and in the $\bar{d}-\bar{u}$ difference.

Equally, the JAM collaboration has presented a new analysis including the latest electroweak boson production data from LHC. These play a crucial role in constraining the up and down sea distributions. This new extraction was focused on exploring the high-$x$ region by lowering the invariant mass cut from the usual $W^{2}\sim 10\text{ GeV}^{2}$ to $W^{2}=3 \text{ GeV}^{2}$. In order to achieve a good description of the high-$x$, low $W^{2}$ data, higher twist corrections, including target mass corrections, were needed. Nuclear effects for the DIS data off deuterium were also considered (see Subsec. \ref{subsec:deuterium}).  
 
The NNPDF collaboration presented the ongoing work towards the NNPDF4.0 set, an update of their previous installment NNPDF3.1 \cite{NNPDF:2017mvq}. The number of points fitted has been extended with the single top, $W+$jet, isolated photon, and di-jets measurements at the LHC. The DIS$+$jet data from HERA were also considered by means of a re-weighting technique. The main differences include, among others, having a parametrisation for the charm quark and its NNLO mass corrections, a new parametrisation, a new optimisation strategy based on gradient descent algorithms and positivity constraints for the gluon and light flavours PDFs\cite{Candido:2020yat}. For the latter there was a dedicated talk on the positivity of the PDFs in the $\bar{MS}$ scheme ($\bf ID=396$). Some exceptions aside, the fit has an overall good quality, with the LHC data playing a crucial role in separating the non-strange sea quarks. However it is noticed that the flavour separation strongly depends on having an intrinsic or a perturbative charm distribution, the former being preferred as it provides a better description of the LHC data. The code used to produce this PDF set is also to be publicly available.  

A different approach to improve the PDFs has been proposed in recent years by the JAM collaboration. Instead of using only inclusive data, they advocate for the simultaneous determination of the collinear PDFs and fragmentation functions (FFs). Here the group presented their results for fitting pion, kaon and charged hadron production and proton PDFs, using a multi-step fitting procedure \cite{Moffat:2021dji}. The inclusion of SIDIS data impacts the sea distributions, in particular that of the strange quark which is suppressed with respect to the case of not having the kaon data in the fit.

In addition to updates from the PDF fitting groups, multiple experiments also presented detailed studies of PDF fits, highlighting their unique contributions to global constraints.  For example, the ATLAS Collaboration presented a new PDF fit~\cite{ATLAS:2021qnl} using $W/Z$ and $W/Z$+jets data from the LHC Run 1 combined with HERA data.  The addition of $V$+jets data significantly reduces the uncertainty and the new fit shows more strangeness suppression than older fits.  High-$x$ studies were presented by the Zeus Collaboration~\cite{ZEUS:2020ddd}.  The LHCb Collaboration discussed their contributions to the quark PDFs at high-$x$ with both their primary $pp$ collision data as well as their fixed-target collision data using the Measuring Overlap with Gas (SMOG) system~\cite{LHCb:2018jry}.

%PDFs
%https://indico.bnl.gov/event/9726/contributions/46570/attachments/33729/54286/boettcher_DIS_14April2021_v2.pdf
%https://indico.bnl.gov/event/9726/contributions/46581/attachments/33747/54308/dis2021-pdf-extr.pdf
%https://indico.bnl.gov/event/9726/contributions/46563/attachments/33664/54180/DIS2021_RA.pdf

\subsection{DIS off heavy-nuclei}\label{subsec:neutrino}

Another issue to be addressed is the problem of disentangling the different parton flavours, which are linked both by their specific combinations in the observables and the scale evolution. This can be done by including in the fits data that are sensitive to other, linearly independent, combinations of PDFs. To this purpose the charged current (CC) DIS process with the exchange of a $W^{+/-}$ boson is an ideal candidate. However most of the data available comes from the collision of neutrino beams with fixed nuclear targets, which not only limits the kinematic coverage of the data but also introduces the problem of correctly quantifying the nuclear effects. Moreover, tension exists between data sets (see Sec. \ref{sec:npdfs}) which hints at the non-universality of nuclear effects.  

In the case of the NNPDF collaboration, they are considering these corrections in NNPDF4.0, in the form of theoretical uncertainties as opposed to the traditional approach of modifying the computed central values using nuclear PDFs or theoretical models. The procedure consists on comparing the values of the nuclear observables computed with and without the nNNPDF2.0 set, and adding them to the fit in the form of a theoretical nuclear covariance matrix \cite{Ball:2020xqw}. The overall result, together with the modifications due to deuterium targets (see Subsec. \ref{subsec:deuterium}), is a significant improvement in the quality of the fit, which highlights the relevance of nuclear effects to the improvement of proton PDFs.

\subsection{DIS with deuterium targets}\label{subsec:deuterium}

A different approach to solve the flavour decomposition problem is to use neutron data, under the form of deuterium beams/targets. While in this case the nuclear effects are smaller than for CC DIS, the fact remains that deuterium is a bound nucleus and, when aiming for a high precision extraction of the proton PDFs, this can not be overlooked. 

The CTEQ and CJ groups have presented a detailed study of the impact of the nuclear corrections in deuterium in their corresponding fits \cite{Accardi:2021ysh}, using the Hessian sensitivity method \cite{Hobbs:2019gob}. In the case of CT18NNLO, the cut on the invariant mass $W$ in DIS restricts the impact of including nuclear modifications, though the description of some observables does improve. By contrast, as the CJ15 fit has a less stringent kinematic cut and thus the sensitivity to the modeled nuclear effects is more striking, with the $\chi^{2}$ decreasing $\sim 5\%$. 

The nuclear modifications for deuterium targets were also newly addressed in the update studies of the JAM and NNPDF collaborations mentioned above. In the case of the JAM collaboration different implementations of the corrections using theoretical models from Nuclear Physics were used, all giving consistently similar fits. The NNPDF collaboration used instead an iterative procedure that permits to determine both the deuteron and proton PDFs, the latter with deuteron uncertainties. While the impact is less significant than the one found for the CC data, they are nonetheless relevant when aiming for higher precision in the proton distributions.

\subsection{Other topics related to proton PDFs}\label{subsec:others}

The significant changes in the analyses of the proton PDF of CT and MSHT have compelled to perform a benchmarking exercise with these two PDF sets and NNPDF3.1. Preliminary results using a reduced data set common to the three PDF extractions, with emphasis on the future precision physics at the LHC, have been presented. With the limited amount of data a good agreement is observed and progress has been made towards a future combination of the individual PDF extractions into a PDF4LHC21 set.

\section{Nuclear PDFs} \label{sec:npdfs}

The correct characterisation of nuclear effects in the parton distributions is important not only due to their relevance in the determination of the proton PDFs and as baseline for new phenomena in heavy-ion collisions. Ultimately, the world around us is composed of nucleons bound together in more complex structures and therefore the nPDFs, that condense in a few parameters the behaviour of partons inside nuclei, are interesting on their own right. 

Dedicated studies of nPDFs have existed for decades now, the extractions done in the same framework as the free proton PDFs, and having them as guidance. The nuclear effects are generally added as a modification of a baseline PDFs, either by expanding the parametrisation or multiplying it by a factor, in both cases with a dependence on the atomic mass number $A$. In this conference three of the groups presented new results: EPPS and NNPDF the status of their updates, nCTEQ specific studies to extend the nCTEQ15 analysis. As before, we will mention only the changes with respect to their previous work.

The EPPS group has extended the data set of EPPS16 \cite{Eskola:2016oht} by including the JLAB NC DIS data, the full CMS di-jets data (replacing the previously used forward-backward ratio), $D^{0}$ meson production from LHCb and the $W$ boson production from CMS in the Run2. All the LHC data considered are from $p+Pb$ collisions. The high precision JLAB data are perfectly described taking into account leading target-mass corrections, with no hints of tensions with the assumption of isospin validity, confirming previous results \cite{Paukkunen:2020rnb}. Regarding the $D^{0}$ production, both the forward and backward data are very well accommodated in the fit. The $W$ boson data is equally well described and does not seem to provide extra constraints with respect to those imposed by the Run1 and CC DIS. As for the di-jets data, the ratio to the $p+p$ reference was fitted. Significant tension was found between the highest forward rapidity bins and all other data in the fit despite implementing a more flexible functional form. Beyond having more data and parameters, the new study includes for the first time the proton uncertainties by fitting the nuclear modifications separately for each error set of their chosen proton PDF baseline. While these are not dominant for most experiments considered and can be reduced by using ratios instead of absolute cross-sections, they become sizable for some observables. The overall effect on the nPDFs is some reduction of the uncertainty bands in the sea sector and, more importantly, a better constraint of the gluon density at high-$x$, pointing to a real anti-shadowing (enhancement) effect.

The NNPDF collaboration has presented the current status of their next instalment of medium modified distributions: nNNPDF3.0. The have now performed a NNLO fit, including $D^{0}$ meson data from LHCb, and electroweak boson production and di-jet data from CMS, in $p+Pb$ collisions at the LHC, the last two using K-factors. Their proton baseline is also updated to the NNPDF4.0 set (see Subsec. \ref{subsec:adddata} for details). As it happens in the EPPS analysis, the extreme rapidity bins of the di-jet data can not be accommodated and a reasonable fit would require their removal. While the use of the missing correlated uncertainties might play a role here, studies performed with nNNPDF2.0 \cite{AbdulKhalek:2020yuc} and their baseline NNPDF3.1 suggest that the main problem is the inability of the latter to describe the $p+p$ di-jet data. Further investigations and the incorporation of the $Z$ and $W$ production data from ATLAS are ongoing.

From their part, the nCTEQ group have presented three studies, each addressing different aspects for improvement. One of them, in collaboration with the CJ group, touches upon the use of available data from NC DIS in the high-$x$ region, in particular the high precision JLAB $6\text{ GeV}$ data, and the impact on the valence quark distributions ({\bf D=452}). By relaxing the cuts on virtuality and invariant mass more data enters in this \emph{nCTEQ15-HIX} fit \cite{Segarra:2020gtj}, in a kinematical region where higher twists effects become relevant. In particular those considered are target-mass corrections and non-perturbative multi-quark interactions. The impact of modelling the nuclear effect in deuterium in the same way as the CJ group is also considered. Fits including only one type of correction or both of them show a significant improvement in the quality of the fit w.r.t. no modifications, with the deuterium dominating the decrease of the $\chi^{2}$. Preliminary results considering data with $x<0.7$ present a change of up to $20\%$ in the $0.6<x<0.7$ region, especially for the up and down densities.

The second topic address by the nCTEQ group was the determination of the strange nPDF. In a previous study where the compatibility of the LHC electroweak boson data in $p+Pb$ collisions with the nCTEQ15 fit was analysed, the parametrisation of the strange distribution was found to be too restrictive. More parameters were needed to accommodate the data into the fit, with a consequent change in shape. The work presented studies in detail the compatibility of the CC DIS data (not included in nCTEQ15), the tensions among the different measurements and the impact on flavour separation \cite{Muzakka:2021uon}. Specifically, the dimuon data from CCFR and NuTev, and the NuTev, CDHSW and Chorus data were considered. It is known that a fraction of the NuTeV cross-section has significant tensions with the other experiments. This is particularly true if the correlated uncertainties are not taken into account. In this work, including the fully correlated uncertainties of the NuTeV data falls short at removing the tensions, which remain for $x\lesssim 0.1$ and $x\sim 0.6$. However the main source of incompatibility, defined with a certain criterion, is the low-$x$ data and their exclusion from the fit gives a consistent result. They do have a significant effect in the shape of the nuclear strange PDF, e.g. $f^{Fe}_{s}$ being a factor two larger at $x\sim 10^{-3}$ when the low-$x$ data are not removed. While the inclusion of the LHC data does indeed reduce the uncertainty of the gluon nPDF, the shape of the strange density remains largely unknown. 

Before the LHC di-jet data became available, the only direct constraint over the gluon density included in the nPDF fits came from single hadron production (SIH) in $d+Au$ collisions at RHIC. For this observable the gluon and quark densities enter at the same order in the calculation and thus are ideal to access the gluon nPDF. The down side of it is, aside from possible final state nuclear effects, that it strongly depends on the set of fragmentation functions (FFs) used to describe how a parton coming out of the hard interaction fragments into the detected hadron. With many FF sets available and new precise data from ALICE, the nCTEQ collaboration presented a detailed analysis of the modifications of nPDFs when including SIH data into the nPDF fits nCTEQ15 and nCTEQ15WZ (with electroweak boson production from LHC). In general, equally good fits can be obtained with all FFs considered, of course the resulting nPDFs being different. The most impact is seen in the gluon and strange distributions, with enhancement for $x\sim 10^{-3}$ \cite{Duwentaster:2021ioo}.

\section{TMD PDFs}

Collinear PDFs accurately describe a large variety of processes involving proton initial states with transverse energy comparable to the momentum transfer.  However, the proton structure is more complicated and a growing number of measurements are sensitive to the transverse momentum of the partons within the proton.  In particular, when the transverse momentum of the outgoing partons is much smaller than the momentum transfer, all the way down to scales of order $\Lambda_\text{QCD}$, collinear PDFs are insufficient for describing the data. This can be remedied with the introduction of transverse momentum dependent PDFs (TMD PDFs), which are of significant interest to the particle and nuclear physics communities.  
While TMD PDFs were a component of many talks, there was a dedicated session about theoretical and phenomenological aspects of these objects, as well as global fits at DIS.  These talks spanned theory developments to global fits, to event simulations using TMD evolution~\cite{Hautmann:2017fcj}.  
For example the first determination of TMD photon densities with the Parton Branching method was presented~\cite{Jung:2021mox}. The photon distribution was generated perturbatively without intrinsic photon component using fits to inclusive DIS HERA data. These TMDs give a good description of the CMS dilepton mass measurements at $\sqrt{s} = 13$ TeV for masses between 15 to 3000 GeV . 

There were also presentations showing dedicated experimental measurements probing TMD PDFs. The COMPASS Collaboration showed new preliminary results for unpolarised Semi-Inclusive DIS of muons on protons. For the TMD physics two observables in unpolarized SIDIS are particularly interesting: azimuthal asymmetries  and transverse momentum distributions, both presented by COMPASS. Both observables were measured over a wide kinematic range of $Q^2$, $x$ and $W$. There is a lot of theoretical work to reproduce the experimental distributions, the dependence on $Q^2$ is expected from the predictions and the behaviour as a function of $W$ can be studied using the COMPASS new results. 

The H1 Collaboration measured the NC jet production as a function of the jet transverse momentum and pseudorapidity, as well as lepton-jet momentum imbalance $q_T$ and azimuthal angle correlation~\cite{h1TMD}. The cross sections were compared to TMD calculations which, without free parameters, described the data over wide kinematic range. In comparison with the collinear predictions, the TMD calculation gives a better description at low $q_T$ and the collinear calculation at large $q_T$. The H1 data can be used to constrain matching between TMD and collinear pQCD frameworks.

Yet another another measurement sensitive to unpolarized TMD PDFs was presented by the STAR Collaboration. Preliminary results on differential $Z$ cross section as a function of the $Z$ transverse momentum were compared to various TMD PDFs. The data are reasonably described by these calculations and shows a possibility to constrain them further.

%Global fits:
%https://indico.bnl.gov/event/9726/contributions/46580/attachments/33871/54522/Scimemi-Spring2021.pdf

%MC methods:
%https://indico.bnl.gov/event/9726/contributions/47419/attachments/33861/54504/DIS2021_LissaKeersmaekers.pdf
%https://indico.bnl.gov/event/9726/contributions/46559/attachments/33845/54481/Talk-DIS-Taheri-15-04-2021.pdf

%Calculations with TMD PDFs
%https://indico.bnl.gov/event/9726/contributions/46564/attachments/33855/54495/sapeta-dis16.pdf

\section{Lattice PDFs}\label{sec:lattice}

The collinear PDFs can be formally defined in terms of matrix elements of specific bi-local operators on the light cone. However, these can not be analytically calculated and we have to recur to global fits such as the ones mentioned in Sec. \ref{sec:pdfs}. In Lattice QCD the computation of parton densities from first principles is also not feasible as long as this definition is used, as the light-cone distances shrink to the origin in Euclidean space-time. Nonetheless, parton densities (collinear, unintegrated, generalised) from the Lattice are starting to become a reality due to several approaches being implemented \cite{Cichy:2018mum}. While there are several options, we focus here in the two that were part of the session. 

On the one hand we have the \emph{pseudo}-PDFs \cite{Radyushkin:2017cyf}. They generalise the PDFs onto spacelike intervals and are defined as the Fourier transforms of the Ioffe-time distributions (ITD) \cite{Ioffe:1969kf,Braun:1994jq}; in the limit of the interval going to zero and with appropriate matching conditions, one recovers the collinear PDFs. This means that if the interval is small its inverse plays a role similar to the renormalization parameter of the scale dependent PDFs (with some differences). One can then write evolution equations for the reduced ITD. The HadStruct collaboration has presented the results needed to calculate the matrix elements and the matching condition for obtaining the gluon PDFs in the pseudo-PDF approach \cite{Balitsky:2019krf}. Using the external field method \cite{Balitsky:1987bk}, the gluon-gluon and gluon-quark kernels were computed. By parametrising the collinear gluon and singlet quark combination, the values of the parametres can be found by fitting to the Lattice results for the matrix elements. 

The same group has also presented the determination of the nucleon valence quark distribution using a pion mass of $m_{\pi}=350$ MeV and lattice spacing $a=0.091$ fm, and applying different techniques for the computation of the matrix elements. The real part of the inverse Fourier transform, which gives the valence distributions, was performed by writing the valence PDF as an Euler Beta function multiplied by a polynomial of degree one, the parameters determined by the Lattice matrix elements. The result is a PDF that closely resembles in shape, but overshoots, the fitted PDFs for $x>0.1$. In the $x<0.1$ region the comparison is significantly worse, though there is no agreement between the PDF sets either. 

On the other hand we have the \emph{quasi}-PDFs, also defined as the Fourier transform of the ITD with respect to an space-like separation that only has one spatial component, in the frame in which the proton momentum is in that particular direction \cite{Ji:2013dva}. The collinear PDFs are recovered from the quasi-PDFs by taking the limit of infinite proton momentum. The Lattice Parton Collaboration presented a self-renormalization method to deal with the divergence that appears in the calculation of the quasi-light-front correlator in the context of the Large Momentum Effective Field Theory (LaMET) \cite{LatticePartonCollaborationLPC:2021xdx}. The main idea of this model independent approach is to perform the extraction of the renormalization factor and residual contribution directly from the matrix element. 

In the contect of quasi-PDFs, the determination of pion valence PDFs and electromagnetic form factors were presented. For the PDFs the calculation was performed using the physical pion mass, which was found to have a limited impact on the distributions. The reconstruction of the x-dependence of the PDFs was done with the the first few moments, extracted using both the NLO and NNLO matching formulas. The latter gave a far better agreement with the pion PDFs obtained from fits than the former. The pion form factors and charge radius were also computed, in good agreement with the measured data. Further improvements are under consideration, including the possible use of a matching formula that incorporates resummation, and the extraction of higher moments.

\section{Experimental Probes}
\label{sec:exp}

%Charged and neutral drell-yan
%https://indico.bnl.gov/event/9726/contributions/46577/attachments/33612/54100/Nam_DIS2021_v11.pdf
%https://indico.bnl.gov/event/9726/contributions/46585/attachments/33841/54542/dis2021.pdf
%https://indico.bnl.gov/event/9726/contributions/46547/attachments/33818/54443/bbilin_CMS_W_Z_DIS2021.pdf

Production of $W$ and $Z$ bosons in proton-proton collisions and fixed target experiments provides stringent tests of calculations based on SM and gives access to quark and antiquark content of the proton. Here especially interesting is the antiquark flavor asymmetry in the proton, measured as the the ratio $\bar{d}/\bar{u}$, where the high $x$ region $x > 0.3$ is of particular importance. The SeaQuest Collaboration measured the Drell-Yan muon pair production using a proton beam at an energy of $120$ GeV and liquid hydrogen and deuterium targets~\cite{SeaQuest:2021zxb}. The cross section ratio of the deuterium and hydrogen data were measured in the wide rage of $x$ for $0.13 < x < 0.45$ and were used as an input to extract the  $\bar{d}/\bar{u}$ ratio in this region. For $0.35 < x < 0.45$ this is a first measurement of the antiquark flavor asymmetry in the proton. In the whole measured $x$ range $\bar{d}/\bar{u} > 1$. This results were compared with various theoretical models. They agree well with the predictions of the meson-baryon model~\cite{Alberg:2017ijg} as well as with the predictions of the statistical parton distributions~\cite{Basso:2015lua}. The STAR experiment at RHIC presented a complementary measurement of the $W$ and $Z/\gamma ^{*}$ differential and total cross sections in p+p collisions at $\sqrt{s} = 500$ GeV and $510$ GeV~\cite{star:2021}. These measurements were taken at high $Q^2$ and $0.1 < x < 0.3$ and can serve as input into global analyses to provide constraints on the sea quark distributions. Another set of measurements important in constraining distributions of the valence and sea quarks in the proton was provided by the CMS Collaborations, obtained in proton-proton collisions at $\sqrt{s} = 13$ TeV. It included measurements of the Drell-Yan processes~\cite{CMS:2018mdl}, $Z$ differential cross sections~\cite{CMS:2019raw} and the study of the $W$ boson rapidity, helicity, double-differential cross sections and charge asymmetry~\cite{CMS:2020cph}. Additionally various rare decay channels of the $W$ were explored, including decays into three pions~\cite{CMS:2019vaj} and a pion and a photon~\cite{CMS:2020oqe}. This wide range of results also allows probing perturbative and non-perturbative QCD effects and rare decay channels, as well as provides valuable input to improve theoretical models. The data are used to test various MC approaches and their validity at different corners of the LHC phase-space.

A complement to dilepton processes is processes involving pairs of quarks and gluons.  While jet cross sections are among the first measurements performed at hadron colliders, we heard important updates from STAR and ATLAS on high-$x$ physics where statistics are limited.  In particular, STAR presented a new measurement of the inclusive dijet cross section, extending previous results with polarized $pp$ collisions~\cite{PhysRevD.95.071103}.  These measurements have important constraint power for the high-$x$ behavior of the gluon PDF.  The ATLAS Collaboration presented a variety of measurements of the $t\bar{t}$ differential cross section.  This included a new result using $\sqrt{s}=5.02$ TeV~\cite{ATLAS-CONF-2021-003}, which provides an interesting data point for comparing with the larger LHC datasets collected at $\sqrt{s}=7,8$, and $13$ TeV.  These data provide a consistent picture across center of mass energies and have constraining power to distinguish between various PDF sets. 

As the namesake of the DIS conference series, it was exciting to also hear updates from high energy lepton-hadron collisions.  The COMPASS experiment at CERN presented a variety of results on semi-inclusive deep inelastic (SIDIS) and deeply virtual Compton scattering.  These data provide powerful insight into azimuthal asymmetries and TMD PDFs.  New data with a proton target are now being analyzed and preliminary results were discussed in the context of the measured energy spectra and angular asymmetries.  The H1 Collaboration also presented a new result sensitive to TMD PDFs, in DIS at higher $Q^2$ with azimuthal decorrelations using jets~\cite{H1prelim-21-031}.  This measurement reports the differential cross section of jets in the Born-level configuration.  One particularlly interestin feature of this new measurement is that it involved the first machine learning-based unfolding~\cite{omnifold} in particle or nuclear physics.  With this technique, spectra can be corrected for detector effects without binning and in a large number of dimensions simultaneously. 

%\textbf{TODO: BN}

%lepton-hadron collisions
%https://indico.bnl.gov/event/9726/contributions/46587/attachments/33853/54513/DIS_2021_Ventura.pdf
%https://indico.bnl.gov/event/9726/contributions/46565/attachments/33866/54509/Moretti_210408_DIS.pdf
%https://indico.bnl.gov/event/9726/contributions/46542/attachments/33840/54474/H1leptonjet_DIS2021.pdf

%jets
%https://indico.bnl.gov/event/9726/contributions/46560/attachments/33898/54571/dis_2021.pdf
%https://indico.bnl.gov/event/9726/contributions/46582/attachments/33795/54387/DIS2021_ttbarDiff_Spalla.pdf

\section{Structure functions beyond PDFs}

Collinear and un-integrated distributions functions are not the only way of describing the measurements. If one is interested in phenomena occurring in regions where pQCD is not fully tested/applicable or its use is inconvenient, other perspectives can become a more suitable choice. In this section we summarise two presentations that take this approach: the first one for photon initiated processes at the LHC, the second one for the study of the DIS structure functions in the low $x$ and $Q^{2}$ region that will be covered at the EIC. 

As discussed in Sec. \ref{sec:pdfs}, the increasing precision at the LHC is reaching the stage at which QED processes become as relevant as the pure QCD ones. For some observables the photon initiated processes play a key role, the LHC effectively acting as a $\gamma+\gamma$ collider. The computation of these contributions can be done in two ways. One approach is to determine the photon content of the proton by means of a photon PDF, to be extracted from fits. This has been and is done with high precision at LO in $\alpha_{s}$, but this feature does not translate well to the calculation of observables, with scale dependence uncertainties washing away the precision of the photon PDF. A second alternative avoids this problem altogether, by not introducing an extra distribution. Instead, the $p\to \gamma+X$ vertex (with $X$ anything) is written in terms of the proton structure function. In the work presented in this session showed the phenomenology of inclusive photon-initiated lepton production \cite{Harland-Lang:2021zvr} and exclusive processes with rapidity gaps, and their implementation in Monte Carlo Event Generators. 

Extracting the DIS one photon exchange cross-section from measurements requires a good understanding of the radiative corrections that alter the kinematics of the initial and final particles with respect to the those entering/leaving the hard interaction. This, in turn, calls for a good theoretical calculation of the structure functions $F_{2}$ and $F_{L}$. The latter, that emerges from the exchange of a longitudinally polarised virtual photon, dominates the low-$x$ region. As in $F_{L}$ the quark and gluon PDFs enter at the same order in the perturbative expansion, measuring it would in principle enable us to study the gluon density, at least for low values of $x$ where it dominates the cross-section. Unfortunately the extraction of $F_{L}$ is inherently non trivial and data for $x<0.01$ are scarce. There are several theoretical models, with different ingredients and assumptions, that aim to describe the structure functions in the low-$x$ region, and the EIC will have the power to discriminate between them. In this line, there was a presentation of a model that extends the structure functions in the experimentally reachable low-$x$, low $Q^{2}$ region. The work is an update of the model proposed in \cite{Badelek:1996ap}, where the extrapolation is based on the photon-gluon fusion process with the quarks masses taken into account and an explicit higher twist effect depending on the transverse momentum of the quarks and the integrated gluon PDF. The use of a more sophisticated model for the higher twist term and recent sets of proton PDFs gives an overall adequate description of the $F_{L}$ data from H1, with the LO PDF giving a better agreement. As seen in the original work, the model severely undershoots the data available at $x>0.01$.

\section*{Acknowledgements - instead of Outlook and Summary}
We would like to thank the DIS conference organizers for the opportunity to chair the WG1 session and all of the speakers and participants for an interesting and stimulating atmosphere. We were originally asked to chair the session at the 2020 iteration of the workshop and we appreciate everyone's patience and understanding as we tried to organize a useful virtual session in 2021 due to the Covid-19 pandemic.

% TODO: include author contributions
%\paragraph{Author contributions}
%This is optional. If desired, contributions should be succinctly described in a single short paragraph, using author initials.

% TODO: include funding information
\paragraph{Funding information}
P. Zurita is supported by the Deutsche Forschungs-gemeinschaft (DFG, German Research Foundation) - Research Unit FOR 2926, grant number 409651613.  B. Nachman is supported by the U.S. Department of En- ergy (DOE), Office of Science under contract DE-AC0205CH11231.
%Authors are required to provide funding information, including relevant agencies and grant numbers with linked author's initials. Correctly-provided data will be linked to funders listed in the \href{https://www.crossref.org/services/funder-registry/}{\sf Fundref registry}.

% TODO:
% Provide your bibliography here. You have two options:

% FIRST OPTION - write your entries here directly, following the example below, including Author(s), Title, Journal Ref. with year in parentheses at the end, followed by the DOI number.

% \bibliographystyle{SciPost_bibstyle} % Include this style file here only if you are not using our template
%\bibliography{SciPost_Example_BiBTeX_File.bib}

\nolinenumbers

\end{document}